\documentclass[letterpaper]{article} 
\usepackage{aaai2026}  
\usepackage{times}  
\usepackage{helvet}  
\usepackage{courier}  
\usepackage[hyphens]{url}  
\usepackage{graphicx} 
\usepackage{natbib}  
\usepackage{caption} 
\usepackage{algorithm}
\usepackage{algorithmic}
\usepackage{amsmath}
\usepackage{amssymb}
\usepackage{multirow}
\usepackage{booktabs}
\usepackage{tabularx}
\usepackage{adjustbox}
\usepackage{makecell}

\usepackage{newfloat}
\usepackage{listings}
\DeclareCaptionStyle{ruled}{labelfont=normalfont,labelsep=colon,strut=off} 
\floatstyle{ruled}
\newfloat{listing}{tb}{lst}{}
\floatname{listing}{Listing}
\nocopyright

\title{Nexus-INR: Diverse Knowledge-guided Arbitrary-Scale Multimodal Medical Image Super-Resolution}
\author{
    Bo Zhang\textsuperscript{\rm 1}, 
    Jianfei Huo\textsuperscript{\rm 1}, 
    Zheng Zhang\textsuperscript{\rm 2}, 
    Wufan Wang\textsuperscript{\rm 1}, 
    Hui Gao\textsuperscript{\rm 3}, 
    Xiyang Gong\textsuperscript{\rm 1}, 
    Wendong Wang\textsuperscript{\rm 1}
}
\affiliations{

    \textsuperscript{\rm 1} State Key Laboratory of Networking and Switching Technology, School of Computer Science (National Pilot Software Engineering School), Beijing University of Posts and Telecommunications.\\
    \textsuperscript{\rm 2} School of Intelligent Engineering and Automation, Beijing University of Posts and Telecommunications.\\
    \textsuperscript{\rm 3} School of Computer Science (National Pilot Software Engineering School), Beijing University of Posts and Telecommunications.\\
    zbo@bupt.edu.cn, jfhuo@bupt.edu.cn, zhangzheng@bupt.edu.cn, wufanwang@bupt.edu.cn, gaohui786@bupt.edu.cn, xygong@bupt.edu.cn, wdwang@bupt.edu.cn
}

\usepackage{bibentry}

\begin{document}

\maketitle

\begin{abstract}

Arbitrary-resolution super-resolution (ARSR) provides crucial flexibility for medical image analysis by adapting to diverse spatial resolutions. However, traditional CNN-based methods are inherently ill-suited for ARSR, as they are typically designed for fixed upsampling factors. While INR-based methods overcome this limitation, they still struggle to effectively process and leverage multi-modal images with varying resolutions and details. 
In this paper, we propose Nexus-INR, a Diverse Knowledge-guided ARSR framework, which employs varied information and downstream tasks to achieve high-quality, adaptive-resolution medical image super-resolution. 
Specifically, Nexus-INR contains three key components. A dual-branch encoder with an auxiliary classification task to effectively disentangle shared anatomical structures and modality-specific features; a knowledge distillation module using cross-modal attention that guides low-resolution modality reconstruction with high-resolution reference, enhanced by self-supervised consistency loss; an integrated segmentation module that embeds anatomical semantics to improve both reconstruction quality and downstream segmentation performance. 
Experiments on the BraTS2020 dataset for both super-resolution and downstream segmentation demonstrate that Nexus-INR outperforms state-of-the-art methods across various metrics.
\end{abstract}

\section{Introduction}

Traditional convolutional neural networks (CNNs) based super-resolution (SR) methods have achieved remarkable progress in medical super-resolution to facilitate downstream analysis, visualization, and quantitative assessment~\cite{chen2018brain, du2020super, 7950500}. Nevertheless, these approaches are typically designed for fixed integer upsampling factors and predefined output grids, making them inflexible for arbitrary-resolution super-resolution (ARSR)~\cite{sitzmann2020implicit, tancik2020fourier, cao2023ciaosr} tasks. 

\begin{figure}[t]
    \centering
    \includegraphics[width=0.95\linewidth]{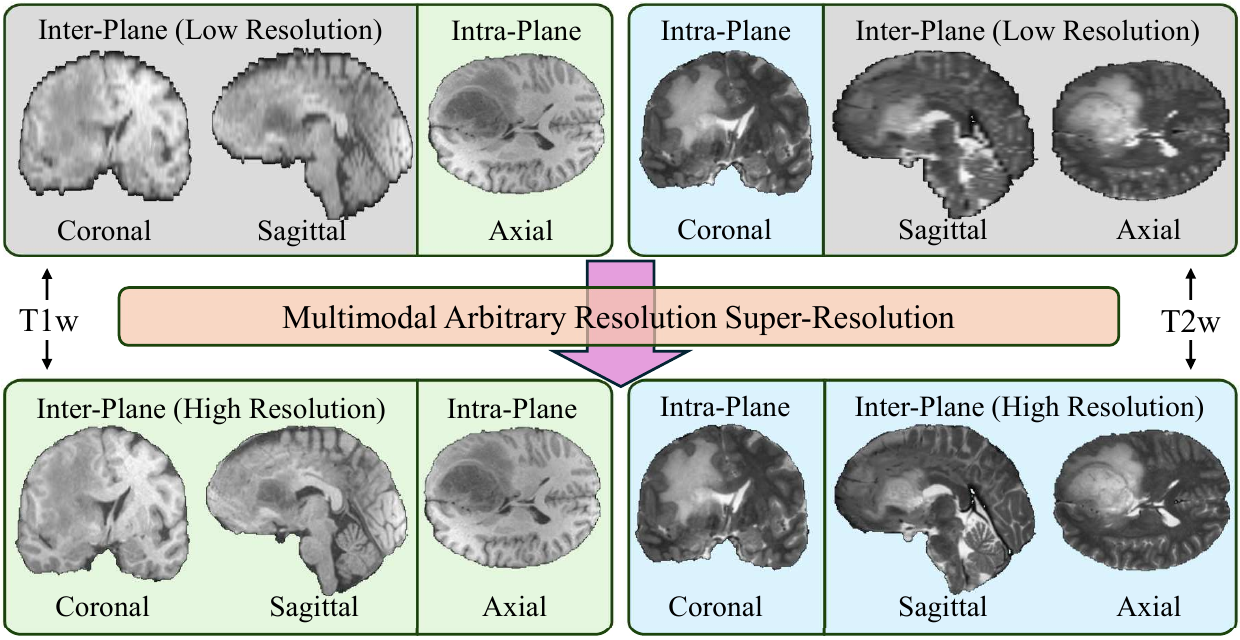}
    \vspace{-2mm}
    \caption{
        Schematic of multi-modal MRI acquisition. Each modality (e.g., T1w and T2w) can have a different intra-plane (scanning) direction (arrows), with high in-plane resolution and low through-plane resolution. T1w images may have higher resolution than T2w images. The slice orientation and resolution anisotropy of each modality lead to cross-modal heterogeneity.
    }
    \label{fig:multimodal_schematic}
    \vspace{-4mm}
\end{figure}

Implicit neural representation (INR) based methods have recently emerged as a promising solution for ARSR. By modeling images as continuous functions that map spatial coordinates to intensities, INR-based methods enable seamless prediction at any desired resolution or spatial grid. This flexibility allows for adaptation to arbitrary, anisotropic, and non-uniform data, which is highly desirable in practical scenarios. However, current INR-based methods are usually tailored for single-modal or isotropic data, lacking the ability to generalize and fully exploit multi-modal information~\cite{mcginnis2023single, chen2021learning, fang2024cycleinr}.

However, multimodal MRI has been widely used in modern medical examinations to capture details of organs and tumor tissues~\cite{lyu2020multi, feng2021multi, CHEN2025103359, wang2023multi}. As shown in Fig.~\ref{fig:multimodal_schematic}, different modalities are frequently acquired with varying orientations, degrees of anisotropy (e.g., high in-plane but low through-plane resolution), and native resolutions—factors driven by scan time, protocol, and hardware limitations. For example, T1-weighted images may have higher spatial resolution than T2-weighted images, and their slice direction is likely to be different. These discrepancies lead to severe spatial heterogeneity, cross-modal misalignment, and resolution inconsistency in the acquired MRI data. 
Current INR-based ARSR methods often inherit from natural image processing algorithms, meaning they don't consider or can't leverage the rich, complementary information across modalities, usually only addressing single-modality-specific issues. This hinders the full potential of multi-modal MRI, with super-resolution results for low-resolution modalities being particularly unsatisfactory. How to effectively accommodate the diverse characteristics of multi-modal imaging still needs to be explored in depth.

Moreover, the core value of medical images lies in their role in diagnostic tasks, such as tumor detection and classification, where preserving fine texture details is crucial~\cite{Xiao2024SemanticSP, HE2023120174}. 
Current super-resolution methods for medical images usually focus on improving image fidelity (e.g., PSNR), overlooking diagnostic accuracy~\cite{Wang_2020_CVPR}. These traditional visual metrics usually fail to fully evaluate the practical utility of the medical images. For instance, 
a super-resolution model that excels in PSNR might negatively or even detrimentally impact downstream tasks if it introduces artifacts or distorts details in small lesions. Therefore, medical image super-resolution should strive to improve both clarity and diagnostic accuracy.

To address these issues, we propose Nexus-INR, a Diverse Knowledge-guided ARSR, which systematically incorporates three types of diverse knowledge into a unified architecture, can better utilize complementary information between different modalities for better medical image super-resolution, thus providing more accurate data for downstream tasks. Our main contributions are as follows:
\begin{itemize}
    \item We propose a Diverse Knowledge-guided ARSR framework that employs varied information and downstream tasks to achieve high-quality, adaptive-resolution medical image super-resolution.
    \item We design a dedicated module to effectively aggregate complementary information from different modalities and orientations, enabling robust feature representation and reconstruction despite spatial heterogeneity. 
    \item We design a cross-modal knowledge distillation module, which transfers anatomical priors from high-resolution reference modalities to low-resolution target modalities, enriching anatomical details and particularly improving SR performance in challenging modalities.
    \item We introduce an auxiliary segmentation task to explicitly guide the network in optimizing anatomical consistency and semantic structures, thereby enhancing image interpretation and downstream task performance. 
\end{itemize}


\section{Related Work}

\textbf{Image Super-Resolution.} 
Recent years have witnessed significant advances in image super-resolution (SR) for medical imaging. Early deep learning-based approaches, such as SRCNN and its 3D extensions \cite{7950500,chen2018brain}, enabled end-to-end mapping from low resolution (LR) to high resolution (HR) images and outperformed traditional interpolation-based techniques. Further improvements were achieved by employing deeper residual networks \cite{du2020super} and GAN-based models \cite{chen2018brain}, which improved reconstruction of fine anatomical details.

To exploit the complementary information in multi-contrast or multi-modal MRI, several dual-branch, attention-based, or transformer-based frameworks have been proposed. Methods such as MCSR \cite{lyu2020multi}, MINet \cite{feng2021multi}, and SANet \cite{feng2024exploring} leverage easy-to-acquire reference modalities to guide the super-resolution of more challenging modalities, achieving notable improvements in image quality. However, these approaches often rely on accurate image registration and may be sensitive to cross-modal misalignment. 
%
Then feature fusion and alignment mechanisms have been proposed. The DANCE framework uses deformable attention and neighborhood aggregation for robust texture transfer and alignment, performing well even with misregistration \cite{CHEN2025103359}.

In addition, INR-based methods have recently gained attention for medical image SR. These approaches model images as continuous functions of spatial coordinates, enabling arbitrary-scale super-resolution and flexible adaptation to anisotropic data \cite{fang2024cycleinr,mcginnis2023single,wu2022arbitrary}. INR-based methods have shown promising results in both single-modal and multi-modal SR tasks.

\begin{figure*}[!t]
    \centering
    \includegraphics[width=0.9\textwidth]{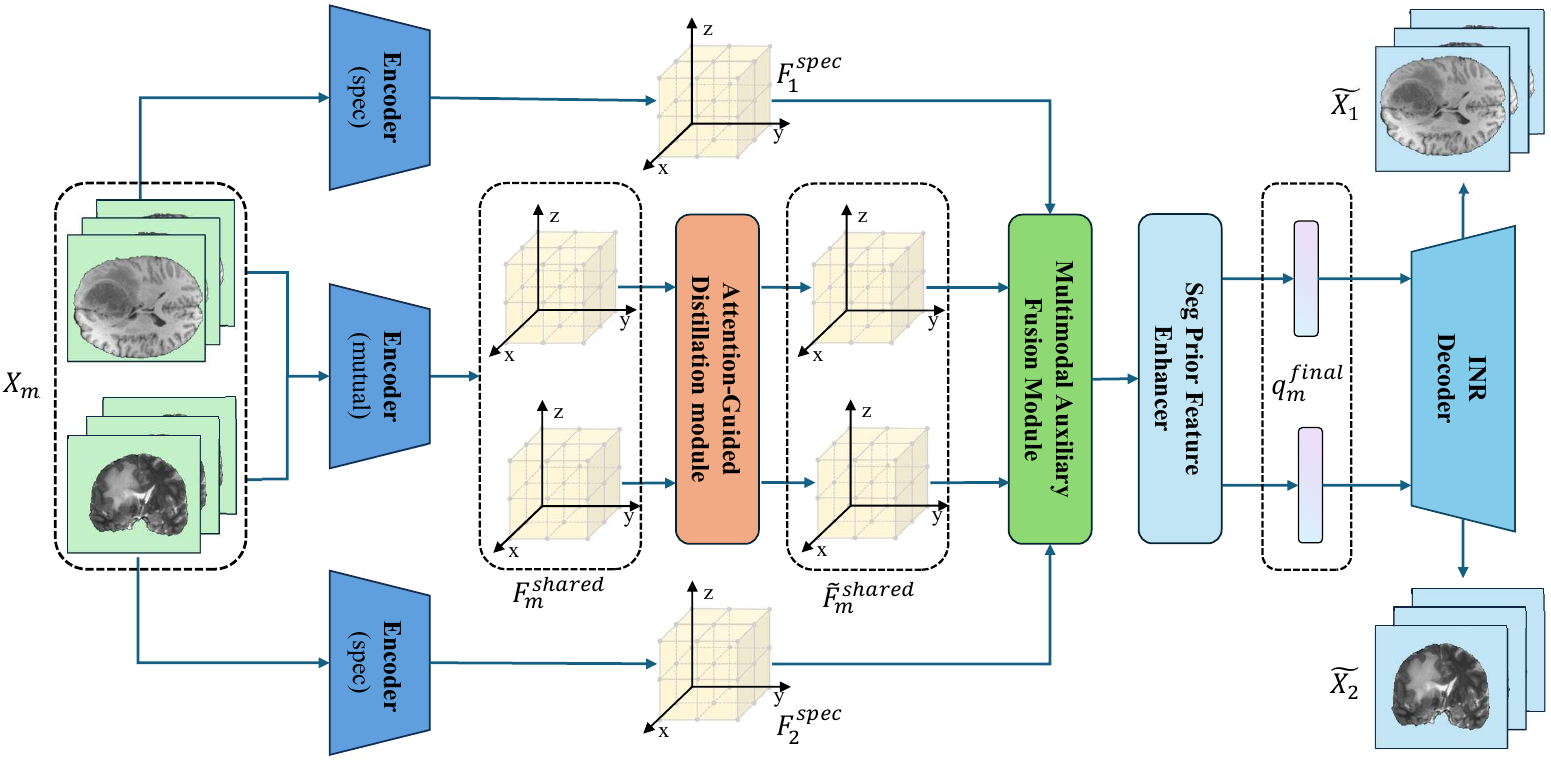}
    \vspace{-2mm}
    \caption{Overview of the Nexus-INR framework. The model consists of dual-branch encoders, cross-modal attention, feature fusion, segmentation enhancement, and a coordinate-based implicit decoder.}
    \label{fig:overview}
    \vspace{-4mm}
\end{figure*}

\textbf{Implicit Neural Representation.} 
Implicit neural representation (INR) methods can 
model images as continuous functions mapping spatial coordinates to pixel intensities. Classic INR approaches represent images by directly fitting a neural network to coordinate-value pairs, enabling voxel-wise prediction at arbitrary resolutions \cite{sitzmann2020implicit, tancik2020fourier}. This 
allows for arbitrary-scale super-resolution and flexible adaptation to diverse spatial grids, which is particularly advantageous in medical imaging SR \cite{wu2022arbitrary, cao2023ciaosr}.

However, the simplicity of directly mapping coordinates to intensities often limits the expressive capacity of basic INRs, especially when modeling complex anatomical structures or generalizing across subjects and modalities. For example, single-subject INR models are typically restricted to fitting a specific image or single subject, and struggle to generalize or transfer to new cases \cite{mcginnis2023single}.

Then various strategies have been proposed to enhance the representational power and generalization ability of INR-based frameworks. One common approach is to incorporate deep features extracted by an encoder, either by concatenating them with spatial coordinates or by using them as conditioning information for the implicit function \cite{chen2021learning, cao2023ciaosr}. Meta-learning methods and transformer-based hyper-networks have been developed to provide strong priors or to directly generate the weights of the INR for a new signal, resulting in faster convergence and improved performance with limited observations \cite{tancik2021learned, chen2022transformers}. Other works introduce semantic or anatomical priors, such as semantic segmentation maps or attention-based mechanisms, to guide the network in capturing both local and non-local features \cite{ekanayake2025seco, cao2023ciaosr}.

The inherent support for ARSR remains a key advantage of INR-based methods, facilitating flexible, memory-efficient, and high-fidelity image reconstruction in both in-scale and out-of-scale scenarios \cite{wu2022arbitrary, cao2023ciaosr, fang2024cycleinr}. With ongoing advances in feature conditioning, meta-learning, transformer-based architectures, and attention modeling, INR frameworks are showing increasing potential for a broad range of medical image super-resolution tasks.

\section{Method}

\subsection{Problem Formulation}
Given a pair of multi-modal MRI volumes, our goal is to reconstruct a high-resolution (HR) image of a target modality from its low-resolution (LR) version, guided by a reference modality. The reference modality is also super-resolved to higher resolution.
Let $\mathbf{X}_1 \in \mathbb{R}^{1 \times D_1 \times H_1 \times W_1}$ and $\mathbf{X}_2 \in \mathbb{R}^{1 \times D_2 \times H_2 \times W_2}$ denote two MRI modalities with possibly anisotropic resolutions, and let $\Omega_{\mathrm{HR}} = \{\mathbf{p}\}$ be the desired HR grid. We aim to learn a continuous implicit function $f_\theta$ parameterized by deep networks:
\begin{equation}
    \hat{y}_m(\mathbf{p}) = f_\theta(\mathbf{X}_1, \mathbf{X}_2, \mathbf{p}),
\end{equation}

\subsection{Overview Framework}

The overall architecture of Nexus-INR is shown in Figure~\ref{fig:overview}. Nexus-INR achieves ARSR for multi-modal MRI by leveraging diverse knowledge from different modalities, anatomical priors, and auxiliary tasks such as segmentation.These knowledge sources are processed through separate branches, each focusing on a specific aspect of the information.

To integrate this diverse knowledge, Nexus-INR extracts features from MRI volumes with anisotropic and mismatched resolutions via parallel encoders, aligns them to a unified HR grid, and applies cross-modal attention for information transfer. Fused features, enhanced by segmentation and positional encodings, are decoded by an implicit network for HR prediction. This approach allows the model to 'diversify' its processing, integrating knowledge at various stages, from feature extraction to reconstruction.

\subsection{Feature Extraction}
Multi-modal MRI provides complementary tissue contrasts but often suffers from inconsistent spatial resolutions due to acquisition protocols. For example, T1w may be acquired at $40 \times 40 \times 20$, while T2w could be $40 \times 10 \times 40$. This spatial heterogeneity presents significant challenges for effective high-fidelity super-resolution.

To address this, our framework explicitly disentangles and exploits both the anatomical structures shared across modalities and the modality-specific details unique. Given paired T1w and T2w volumes $\mathbf{X}_1$ and $\mathbf{X}_2$ (each of shape $[B, 1, D, H, W]$), we employ two parallel encoders for each modality:
1) \textbf{Shared encoder} ($\mathrm{Encoder}_m^{\mathrm{shared}}$): Extracts features representing anatomical structures common to both modalities, enabling robust cross-modal guidance.
2) \textbf{Modality-specific encoder} ($\mathrm{Encoder}_m^{\mathrm{spec}}$): Captures unique contrasts and details essential for preserving modality-specific information during fusion and reconstruction.

Formally, for $m \in \{1,2\}$ (T1w and T2w), the extracted features are
\begin{align}
    \mathbf{F}_m^{\mathrm{shared}} &= \mathrm{Encoder}_m^{\mathrm{shared}}(\mathbf{X}_m), \\
    \mathbf{F}_m^{\mathrm{spec}} &= \mathrm{Encoder}_m^{\mathrm{spec}}(\mathbf{X}_m).
\end{align}
Due to the mismatch in spatial resolutions, all feature maps are upsampled to a unified high-resolution grid (e.g., $40 \times 40 \times 40$) using trilinear interpolation:
\begin{align}
    \tilde{\mathbf{F}}_m^{\mathrm{shared}} &= \mathrm{Interp}(\mathbf{F}_m^{\mathrm{shared}}), \\
    \tilde{\mathbf{F}}_m^{\mathrm{spec}}   &= \mathrm{Interp}(\mathbf{F}_m^{\mathrm{spec}}).
\end{align}
This upsampling ensures spatial alignment for subsequent attention and loss computations, enabling accurate voxel-wise information transfer.

\subsection{Knowledge Distillation via Cross-Modal Attention}
A core challenge in multi-modal super-resolution is how to effectively transfer the rich anatomical priors of the high-resolution modality (T1w) to guide the reconstruction of the low-resolution modality (T2w). To achieve this, we introduce a cross-modal attention mechanism that enables T2w’s shared features to dynamically attend to T1w’s shared features at each location in the unified high-resolution space.

This design is motivated by the observation that simple concatenation or addition of features is insufficient for capturing complex, spatially-varying correlations between modalities, especially under resolution mismatch. Cross-attention allows the model to learn spatially-adaptive, content-aware guidance from the teacher (T1w) to the student (T2w), enhancing the transfer of fine anatomical structures.

\begin{figure}[!t]
    \centering
    \includegraphics[width=0.88\linewidth]{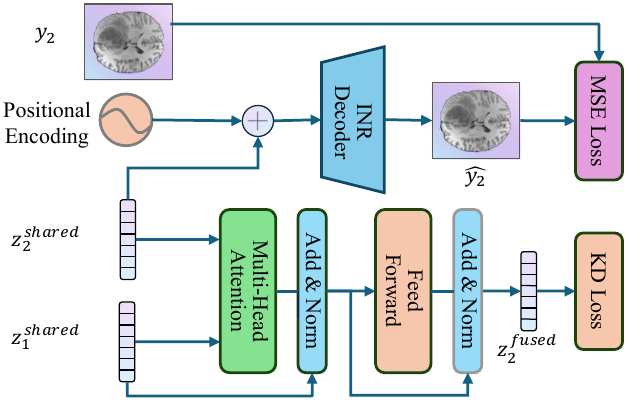}
    \vspace{-2mm}
    \caption{Illustration of the cross-modal attention-based knowledge distillation and T2w self-reconstruction loss. $y_2$ denotes the ground-truth T2w image, while $\hat{y}_2$ denotes the predicted T2w image.}
    \label{fig:kd_t2self}
    \vspace{-3mm}
\end{figure}

The attention-enhanced T2w shared feature is computed as:
\begin{equation}
    \tilde{\mathbf{F}}_2^{\mathrm{fused}} = \mathrm{CrossAttn}(\tilde{\mathbf{F}}_2^{\mathrm{shared}}, \tilde{\mathbf{F}}_1^{\mathrm{shared}}),
\end{equation}
where $\mathrm{CrossAttn}(\cdot, \cdot)$ denotes a cross-attention block with T2w as query and T1w as key/value, operating voxel-wise after upsampling.

To facilitate supervision and ensure that the student branch (T2w) learns to mimic the teacher's (T1w) anatomical encoding, we sample both the T1w shared features and the attention-fused T2w features at a set of high-resolution coordinates $\{\mathbf{p}_i\}_{i=1}^K$:
\begin{equation}
    \mathbf{z}_1^{\mathrm{shared}} = \mathrm{GridSample}(\tilde{\mathbf{F}}_1^{\mathrm{shared}}, \{\mathbf{p}_i\}),
\end{equation}
\begin{equation}
    \mathbf{z}_2^{\mathrm{fused}} = \mathrm{GridSample}(\tilde{\mathbf{F}}_2^{\mathrm{fused}}, \{\mathbf{p}_i\}),
\end{equation}
where $\mathrm{GridSample}(\cdot, \cdot)$ denotes trilinear interpolation and $\mathbf{z}_1^{\mathrm{shared}}, \mathbf{z}_2^{\mathrm{fused}} \in \mathbb{R}^{B \times K \times C}$.

\begin{figure}[!t]
    \centering
    \includegraphics[width=0.88\linewidth]{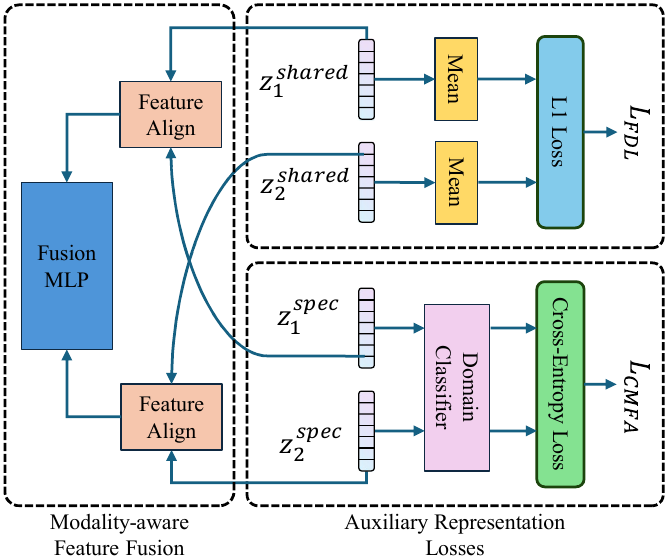}
    \vspace{-2mm}
    \caption{Workflow of auxiliary losses: CMFA aligns the mean of shared features across modalities via L1 loss, while FDL uses a classifier and cross-entropy loss to enforce modality discrimination on specific features.}
    \label{fig:cmfa_fdl}
    \vspace{-3mm}
\end{figure}

Based on the above, we establish a voxel-level knowledge distillation loss:
\begin{equation}
    \mathcal{L}_{\mathrm{KD}} = \frac{1}{B K} \sum_{b=1}^B \sum_{k=1}^K \left\| \mathbf{z}_2^{\mathrm{fused}, (b,k)} - \mathbf{z}_1^{\mathrm{shared}, (b,k)} \right\|_2^2.
\end{equation}
This loss enforces that, at every spatial location, the T2w branch captures the anatomical structures encoded by T1w, thus enabling robust cross-modal guidance for ARSR.

In addition, to further enhance the representation learning for the target modality, we introduce a self-reconstruction loss for T2w. Specifically, the upsampled shared T2w features are sampled at the same set of coordinates and combined with positional encoding, then decoded to predict the high-resolution T2w intensity:
\begin{equation}
    \mathbf{z}_2^{\mathrm{shared}} = \mathrm{GridSample}(\tilde{\mathbf{F}}_2^{\mathrm{shared}}, \{\mathbf{p}_i\}).
\end{equation}
The decoder predicts the intensity $\hat{y}_2^{\mathrm{self}}$ and the loss is defined as:
\begin{equation}
    \mathcal{L}_{\mathrm{T2Self}} = \frac{1}{B K} \sum_{b=1}^B \sum_{k=1}^K \left\| \hat{y}_2^{\mathrm{self},(b,k)} - y_2^{(b,k)} \right\|_2^2,
\end{equation}
where $y_2^{(b,k)}$ is the ground-truth T2w intensity at location $\mathbf{p}_i$.

The total objective for cross-modal knowledge transfer is a weighted sum of the two losses:
\begin{equation}
    \mathcal{L}_{\mathrm{KD\_total}} = \lambda_{\mathrm{KD}} \mathcal{L}_{\mathrm{KD}} + \lambda_{\mathrm{T2Self}} \mathcal{L}_{\mathrm{T2Self}},
\end{equation}
where $\lambda_{\mathrm{KD}}$ and $\lambda_{\mathrm{T2Self}}$ are hyperparameters balancing the two terms.

By jointly optimizing these objectives, our framework enables the T2w branch to effectively absorb high-resolution anatomical priors from the T1w and maintain self-consistency, even in the presence of cross-modal misalignment, thereby significantly improving the quality and anatomical fidelity of the super-resolved outputs.

\subsection{Multi-Modal Feature Fusion by Auxiliary Losses}
After attention-based knowledge transfer, we further integrate the shared and modality-specific features of each modality to obtain discriminative, information-rich representations for downstream reconstruction.

\textbf{Residual Feature Fusion.}
For each modality $m \in \{1,2\}$, we sample both the upsampled shared and specific features at the target coordinates:
\begin{equation}
    \mathbf{z}_m^{\mathrm{shared}} = 
    \begin{cases}
        \mathrm{GridSample}(\tilde{\mathbf{F}}_1^{\mathrm{shared}}, \{\mathbf{p}_i\}), & m=1. \\
        \mathrm{GridSample}(\tilde{\mathbf{F}}_2^{\mathrm{fused}}, \{\mathbf{p}_i\}), & m=2.
    \end{cases}
\end{equation}
\begin{equation}
    \mathbf{z}_m^{\mathrm{spec}} = \mathrm{GridSample}(\tilde{\mathbf{F}}_m^{\mathrm{spec}}, \{\mathbf{p}_i\}).
\end{equation}
We then concatenate the sampled shared and specific features and project them through a multi-layer perceptron (MLP) with a residual connection:
\begin{equation}
    \mathbf{h}_m = [\mathbf{z}_m^{\mathrm{shared}}, \mathbf{z}_m^{\mathrm{spec}}].
\end{equation}
\begin{equation}
    \mathbf{f}_m = f_{\mathrm{proj}}(\mathbf{h}_m) + \mathbf{z}_m^{\mathrm{shared}}.
\end{equation}
Here, the residual design preserves the core anatomical structure while allowing the network to flexibly integrate modality-specific cues.

\textbf{Positional Encoding.}
To provide explicit spatial context, we compute positional encodings (Fourier features) for all target coordinates and concatenate them with the fused features:
\begin{equation}
    \mathbf{q}_m = [\mathbf{f}_m, \mathbf{e}_m],
\end{equation}
where $\mathbf{e}_m \in \mathbb{R}^{K \times d_e}$ is the positional encoding for modality $m$.

\textbf{Auxiliary Losses.}
To further enhance feature consistency and discrimination, we introduce two auxiliary losses, as illustrated in Figure~\ref{fig:cmfa_fdl}:

\textit{Cross-Modal Mean Feature Alignment Loss (CMFA).}
This loss encourages the mean shared features of both modalities to be close, promoting robust cross-modal alignment:
\begin{equation}
    \mathcal{L}_{\mathrm{CMFA}} = \left\| \mathrm{Mean}(\mathbf{z}_1^{\mathrm{shared}}) - \mathrm{Mean}(\mathbf{z}_2^{\mathrm{shared}}) \right\|_1.
\end{equation}
\textit{Feature Discrimination Loss (FDL, Cross-Entropy).}
This loss encourages the modality-specific features to be discriminative, facilitating downstream tasks such as modality classification and robust fusion:
\begin{equation}
    \mathcal{L}_{\mathrm{FDL}} = \mathrm{CE}(f_{\mathrm{cls}}(\mathbf{z}_1^{\mathrm{spec}}), 0) + \mathrm{CE}(f_{\mathrm{cls}}(\mathbf{z}_2^{\mathrm{spec}}), 1),
\end{equation}
where $f_{\mathrm{cls}}$ is a lightweight classifier and $0,1$ are the modality labels for T1w and T2w, respectively.

The final representation $\mathbf{q}_m$ is then ready for implicit decoding to predict high-resolution voxel intensities (details omitted here).

\begin{figure}[!t]
    \centering
    \includegraphics[width=0.88\linewidth]{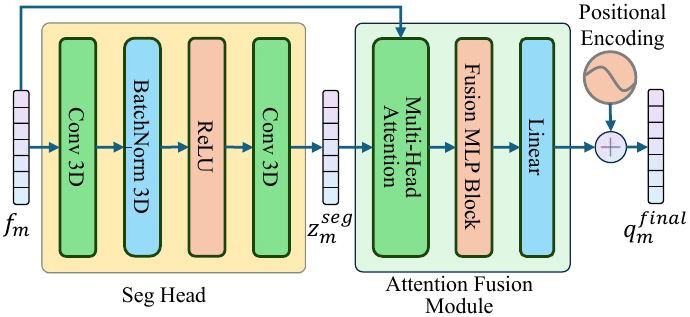}
    \vspace{-3mm}
    \caption{
        Workflow of segmentation-enhanced feature fusion and final implicit decoding. 
    }
    \label{fig:seg_fusion_pe}
    \vspace{-3mm}
\end{figure}

\subsection{Segmentation-Enhanced Learning}

Our framework adopts a multi-task learning strategy, jointly optimizing super-resolution and segmentation tasks with shared encoders. This joint training encourages the learned representations to be informative for both high-fidelity reconstruction and accurate anatomical delineation.

\begin{table*}[!t]
\centering
\small
\begin{tabular}{lccc|ccc|ccc}
\toprule
\multirow{2}{*}{Method} & \multicolumn{3}{c|}{PSNR $\uparrow$} & \multicolumn{3}{c|}{SSIM $\uparrow$} & \multicolumn{3}{c}{LPIPS $\downarrow$} \\
                        & T1w & T2w & Avg & T1w & T2w & Avg & T1w & T2w & Avg \\
\midrule
Trilinear         & 26.50 & 25.10 & 25.80 & 0.8700 & 0.8550 & 0.8625 & 0.0420 & 0.0480 & 0.0450 \\
SMORE~\cite{9253710}             & 29.20 & 27.80 & 28.50 & 0.9400 & 0.9250 & 0.9325 & 0.0280 & 0.0320 & 0.0300 \\
I3Net~\cite{10508991}           & 30.60 & 29.20 & 29.90 & 0.9550 & 0.9400 & 0.9475 & 0.0170 & 0.0210 & 0.0190 \\
SAINT~\cite{peng2020saint}              & \underline{31.50} & \underline{30.10} & \underline{30.80} & 0.9520 & 0.9370 & 0.9445 & 0.0310 & 0.0350 & 0.0330 \\
ArSSR~\cite{wu2022arbitrary}         & 30.01 & 28.58 & 29.30 & 0.9378 & 0.9318 & 0.9348 & \underline{0.0138} & \underline{0.0185} & \underline{0.0161} \\
SSMCSR$^\#$~\cite{mcginnis2023single}       & \underline{31.50} & 30.00 & 30.75 & \underline{0.9630} & \underline{0.9480} & \underline{0.9555} & 0.0280 & 0.0320 & 0.0300 \\
CycleINR$^*$~\cite{fang2024cycleinr}  & 30.60 & 29.20 & 29.90 & 0.9490 & 0.9340 & 0.9415 & 0.0150 & 0.0190 & 0.0170 \\
\textbf{Nexus-INR (Ours)}  & \textbf{33.75} & \textbf{32.25} & \textbf{33.00} & \textbf{0.9772} & \textbf{0.9566} & \textbf{0.9669} & \textbf{0.0038} & \textbf{0.0095} & \textbf{0.0067} \\
\bottomrule
\end{tabular}
    \vspace{-2mm}
\caption{Comparison with state-of-the-art methods on the BraTS2020 test set. The best two results are highlighted in bold and underline. $^*$: CycleINR results are from our own implementation according to the original paper due to unavailable code. $^\#$: SSMCSR is subject-specific and cannot be trained/evaluated on the whole dataset. }
\label{tab:sota}
\end{table*}

\begin{figure*}[!t]
    \centering
    \includegraphics[width=\textwidth]{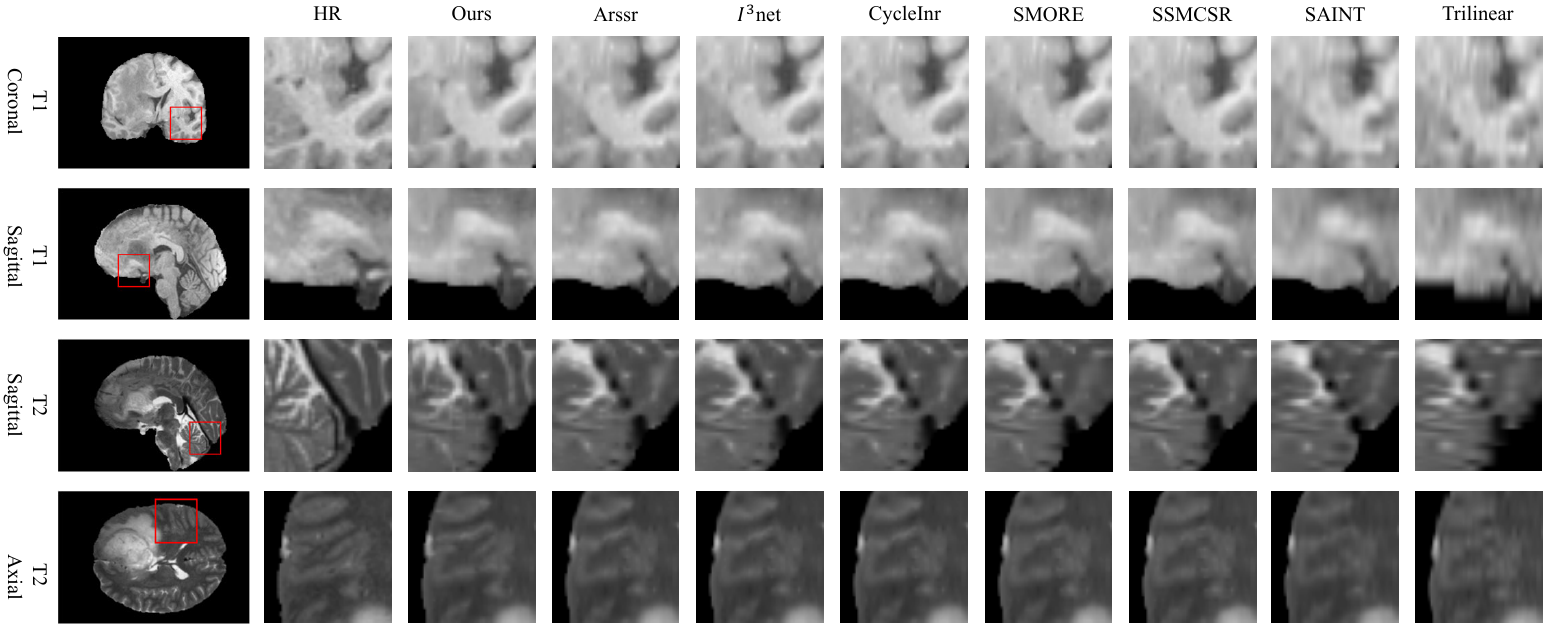}
    \vspace{-6mm}
    \caption{Visual comparison of super-resolved slices produced by different methods on the BraTS2020 test set. 
        Our Nexus-INR achieves sharper boundaries and better anatomical fidelity compared to other state-of-the-art approaches. }
    \label{fig:visual_results}
    \vspace{-3mm}
\end{figure*}

For each modality, the fused features $\mathbf{f}_1$ and $\mathbf{f}_2$ at query coordinates are obtained by integrating shared and specific features via residual MLP fusion, and then passed through segmentation heads to generate segmentation logits:
\begin{align}
    \mathbf{z}_1^{\mathrm{seg}} &= f_{\mathrm{seg}}^1(\mathbf{f}_1), \\
    \mathbf{z}_2^{\mathrm{seg}} &= f_{\mathrm{seg}}^2(\mathbf{f}_2),
\end{align}
where $f_{\mathrm{seg}}^1$ and $f_{\mathrm{seg}}^2$ denote the segmentation heads for T1w and T2w, respectively.

To further enhance super-resolution, segmentation features are fused back with the original features by channel-wise concatenation and a fusion MLP:
\begin{align}
    \mathbf{h}_1^{\mathrm{final}} &= f_{\mathrm{fusion}}^1([\mathbf{f}_1, \mathbf{z}_1^{\mathrm{seg}}]), \\
    \mathbf{h}_2^{\mathrm{final}} &= f_{\mathrm{fusion}}^2([\mathbf{f}_2, \mathbf{z}_2^{\mathrm{seg}}]),
\end{align}
where $f_{\mathrm{fusion}}^1$ and $f_{\mathrm{fusion}}^2$ are fusion MLPs for T1w and T2w, respectively.

Finally, the final features are concatenated with positional encodings for the implicit decoder:
\begin{align}
    \mathbf{q}_1^{\mathrm{final}} &= [\mathbf{h}_1^{\mathrm{final}}, \mathbf{e}_1], \\
    \mathbf{q}_2^{\mathrm{final}} &= [\mathbf{h}_2^{\mathrm{final}}, \mathbf{e}_2],
\end{align}
where $\mathbf{e}_1$ and $\mathbf{e}_2$ are positional encodings for T1w and T2w, respectively. The resulting features are then fed into the implicit decoder to predict high-resolution voxel intensities.

The segmentation heads are supervised by a cross-entropy loss with respect to the ground-truth segmentation masks. The overall objective is a weighted sum of the super-resolution loss, segmentation loss, and auxiliary losses:
\begin{equation}
    \mathcal{L} = \mathcal{L}_{\mathrm{SR}} + \lambda_{\mathrm{seg}} \mathcal{L}_{\mathrm{seg}} + \lambda_{\mathrm{aux}} \mathcal{L}_{\mathrm{aux}},
\end{equation}
where $\lambda_{\mathrm{seg}}$ and $\lambda_{\mathrm{aux}}$ are weighting factors.

This design enables the model to leverage both dense image-level and anatomical supervision, leading to improved reconstruction fidelity, anatomical accuracy, and generalization.

\begin{table*}[t]
\centering
\small

\begin{tabular}{cccc|ccc|ccc|ccc}
\toprule
\multicolumn{4}{c|}{Components} & \multicolumn{3}{c|}{PSNR $\uparrow$} & \multicolumn{3}{c|}{SSIM $\uparrow$} & \multicolumn{3}{c}{LPIPS $\downarrow$} \\
\cmidrule(lr){1-4} \cmidrule(lr){5-7} \cmidrule(lr){8-10} \cmidrule(lr){11-13}
Baseline & MMF & KD-Attn & SEL & T1w & T2w & Avg & T1w & T2w & Avg & T1w & T2w & Avg \\
\midrule
$\checkmark$     &        &     &             & 30.01 & 28.58 & 29.30 & 0.9378 & 0.9318 & 0.9348 & 0.0138 & 0.0185 & 0.0161 \\
$\checkmark$    & $\checkmark$       &     &             & 33.72 & 29.71 & 31.72 & 0.9769 & 0.9324 & 0.9547 & 0.0040 & 0.0207 & 0.0124 \\
$\checkmark$    & $\checkmark$      &  $\checkmark$   &             & 33.63 & 32.07 & 32.85 & 0.9764 & 0.9551 & 0.9657 & 0.0039 & 0.0111 & 0.0075 \\
$\checkmark$    & $\checkmark$      & $\checkmark$   &  $\checkmark$           & \textbf{33.75} & \textbf{32.25} & \textbf{33.00} & \textbf{0.9772} & \textbf{0.9566} & \textbf{0.9669} & \textbf{0.0038} & \textbf{0.0095} & \textbf{0.0067} \\
\bottomrule
\end{tabular}
    \vspace{-2mm}
\caption{Ablation study of Nexus-INR components on BraTS2020 (Nexus-INR = MMF + KD-Attn + SEL). MMF: Multi-Modal Fusion, KD-Attn: Cross-Modal Attention, SEL: Segmentation-Enhanced Learning.}
\label{tab:ablation}
    \vspace{-3mm}
\end{table*}

\begin{table}[t]
\centering
\small
\begin{tabular}{lccc}
\toprule
Encoder & PSNR $\uparrow$ & SSIM $\uparrow$ & LPIPS $\downarrow$ \\
\midrule
SRResNet    & 33.12          & 0.9635          & 0.0084          \\
ResCNN      & 32.85          & 0.9612          & 0.0091          \\
RDN         & \textbf{33.75} & \textbf{0.9669} & \textbf{0.0067} \\
\bottomrule
\end{tabular}
    \vspace{-2mm}
\caption{Comparison of different encoder backbones in Nexus-INR on the BraTS2020 test set.}
\label{tab:encoder}
    \vspace{-2mm}
\end{table}

\begin{table}[t]
\centering
\small
\begin{tabular}{lccc}
\toprule
INR Decoder & PSNR $\uparrow$ & SSIM $\uparrow$ & LPIPS $\downarrow$ \\
\midrule
SIREN      & 33.21          & 0.9639          & 0.0081          \\
MLP        & \textbf{33.75} & \textbf{0.9669} & \textbf{0.0067} \\
\bottomrule
\end{tabular}
    \vspace{-2mm}
\caption{Comparison of different INR decoders in Nexus-INR on the BraTS2020 test set.}
\label{tab:inr}
    \vspace{-3mm}
\end{table}

\begin{table}[t]
\centering
\small
\begin{tabular}{lcccc}
\toprule
\multirow{2}{*}{Method} & \multicolumn{2}{c}{U-Net} & \multicolumn{2}{c}{TransUNet} \\
\cmidrule(lr){2-3} \cmidrule(lr){4-5}
 & Dice $\uparrow$ & HD95 $\downarrow$ & Dice $\uparrow$ & HD95 $\downarrow$ \\
\midrule
Original LR                      & 78.80 & 8.65 & 79.40 & 8.33 \\
Trilinear                        & 82.65 & 7.24 & 83.11 & 7.05 \\
SMORE          & 85.72 & 6.58 & 86.09 & 6.31 \\
I3Net           & 86.18 & 6.31 & 86.73 & 6.12 \\
SAINT      & 86.35 & 6.24 & 86.98 & 6.05 \\
ArSSR    & 87.02 & 5.90 & 87.41 & 5.64 \\
SSMCSR$^\#$ & 87.18 & 5.84 & 87.52 & 5.61 \\
CycleINR$^*$  & 87.24 & 5.79 & 87.63 & 5.55 \\
\textbf{Nexus-INR (Ours)}        & \textbf{88.92} & \textbf{5.03} & \textbf{88.97} & \textbf{4.82} \\
Original HR                      & 90.12 & 4.32 & 91.05 & 3.97 \\
\bottomrule
\end{tabular}
    \vspace{-2mm}
\caption{
Segmentation performance (Dice $\uparrow$, HD95 $\downarrow$) of U-Net and TransUNet on super-resolved images from different methods. 
$^*$: CycleINR results are from our own implementation according to the original paper due to unavailable code. 
$^\#$: SSMCSR is subject-specific and cannot be trained/evaluated on the whole dataset.
}
\label{tab:seg}
    \vspace{-3mm}
\end{table}

\section{Experiments}

\subsection{Datasets and Evaluation Metrics}
To better simulate clinical MRI acquisition, we constructed our dataset from the BraTS collection. Due to T1w images are often acquired at higher spatial resolution than T2w images, and MRI data typically exhibit anisotropic resolution. We therefore downsampled T1w images along the z-axis to $240 \times 240 \times 80$ and T2w images along the y-axis to $240 \times 60 \times 160$, reflecting clinical acquisition protocols with different orientations for each modality. 
A total of 73 subjects with isotropic ground-truth scans were selected, split into 48 for training, 10 for validation, and 15 for testing. All images were normalized and zero-padded to $240 \times 240 \times 160$. T1w and T2w volumes were cropped into $40 \times 40 \times 20$ and $40 \times 10 \times 40$ patches, respectively, and only patches containing sufficient foreground were retained.

For evaluation, we report Peak Signal-to-Noise Ratio (PSNR), Structural Similarity Index Measure (SSIM), and Learned Perceptual Image Patch Similarity (LPIPS) to assess both fidelity and perceptual quality.

\subsection{Implementation Details}
All experiments are implemented in PyTorch and conducted on two NVIDIA RTX 4090 GPUs. The encoder adopts the RDN \cite{Zhang_Tian_Kong_Zhong_Fu_2018} architecture (feature dimension 128), and the implicit decoder is a 4-layer MLP with 256 units per layer. We use the Adam optimizer with a learning rate of $1 \times 10^{-4}$ and batch size 22. The total loss is a weighted sum of all task losses.

MRI volumes are normalized to zero mean and unit variance. Data augmentation includes random flipping and rotation. Model selection is based on the best validation PSNR, and all experiments are repeated with three random seeds for reproducibility. During inference, the model supports arbitrary-resolution prediction by querying the implicit decoder with any desired coordinate grid.

\subsection{Comparison with State-of-the-art Methods}
We compare our Nexus-INR with several state-of-the-art (SOTA) methods on the BraTS2020 test set, where SMORE~\cite{9253710}, I3Net~\cite{10508991}, and SAINT~\cite{peng2020saint} are CNN-based methods, while ArSSR~\cite{wu2022arbitrary}, SSMCSR~\cite{mcginnis2023single}, and CycleINR~\cite{fang2024cycleinr} are INR-based methods. It should be noted that the results of CycleINR are from our own implementation according to the original paper, and the results of SSMCSR is subject-specific. Results of other methods are tested and calculated by ourselves. 

As presented in Table~\ref{tab:sota}, our Nexus-INR achieves the best performance on all metrics. Especially on the lower-resolution T2w modality, it outperforms all others by at least 2.15 PSNR points and leads in SSIM. It also significantly reduces LPIPS, showing remarkable gains in perceptual quality and detail recovery. 
Figure~\ref{fig:visual_results} shows representative super-resolved slices of T1w and T2w modalities by different methods. We can observe that our Nexus-INR more faithfully restores brain structures and fine details, such as tissue boundaries and cortical features, especially in challenging T2w images. 
This shows our method is superior to other algorithms in both qualitative and quantitative evaluations of output images.

\subsection{Ablation Analysis}
We conduct ablation experiments by incrementally adding MMF, KD-Attn, and SEL modules to the baseline model, retraining under the same settings. 
As presented in Table~\ref{tab:ablation}, we can observe that each module incrementally improves PSNR, SSIM, and LPIPS, with KD-Attn notably boosting T2w performance. The full model achieves the best results, confirming the effectiveness of our design.

\subsection{Impact of Encoder and Implicit Representation Choices}
To further evaluate the flexibility and generalizability of our framework, we conduct additional experiments on the BraTS2020 test set by replacing the backbone encoder and INR decoder within Nexus-INR. For the encoder comparison, we evaluate three representative backbones: RDN, SRResNet \cite{SRRESNET}, and ResCNN \cite{Du_He_Wang_Gholipour_Zhou_Chen_Jia_2020}. For the INR decoder comparison, we consider two popular implicit decoders: MLP and SIREN. All other experimental settings are kept consistent with the main experiments.
%
As presented in Table~\ref{tab:encoder} and Table~\ref{tab:inr}, we can observe that our Nexus-INR demonstrates robust performance across different encoder and INR decoder choices, further verifying its versatility and compatibility with a variety of network designs.

\subsection{Impact on Segmentation Task}

To assess the effect of different super-resolution methods on downstream segmentation, we use a unified evaluation pipeline: first, each method is used to generate super-resolved images from low-resolution inputs; then, both a 3D U-Net and a TransUNet~\cite{chen2021transunet} are tested on these images for tumor segmentation. All segmentation models are tested under the same conditions to ensure a fair comparison. 
Segmentation performance is evaluated using the Dice similarity coefficient (Dice, higher is better) and the 95th percentile Hausdorff Distance (HD95, lower is better) on the BraTS2020 test set.
As presented in Table~\ref{tab:seg}, we can observe that our Nexus-INR achieves the best segmentation accuracy and boundary quality for both U-Net and TransUNet, demonstrating that our 
approach is also the most meaningful for downstream tasks.

\section*{Conclusion}

In this paper, we propose Nexus-INR, a dual branch network for high-quality, adaptive-resolution medical image super-resolution. Nexus-INR integrates multi-modal fusion, cross-modal knowledge distillation, and segmentation-enhanced learning to leverage diverse knowledge to handle multi-modal images and provide more accurate data for downstream tasks. Experiments on the BraTS2020 dataset show that Nexus-INR outperforms state-of-the-art methods in both image quality and downstream segmentation performance. 

\bibliography{aaai2026}

\end{document}